\documentclass[12pt]{article}
\usepackage{graphicx,a4} 
\usepackage{amssymb}
\usepackage{latexsym}
\begin{document}

\newcommand{\preprintno}[1]
{\vspace{-2cm}{\normalsize\begin{flushright}#1\end{flushright}}\vspace{1cm}}

\title{\preprintno{{\bf ULB-TH/02-19}}Randall-Sundrum black holes and strange stars}
\author{{Malcolm Fairbairn\thanks{E-mail: mfairbai@ulb.ac.be}} and
{V\'eronique Van Elewyck\thanks{E-mail: vvelewyc@ulb.ac.be}}\\
{\em Service de Physique Th\'eorique, CP225}\\
{\em Universit\'e Libre de Bruxelles, B-1050 Brussels, Belgium}}

\date{March 2003}

\maketitle
\begin{abstract}
\noindent
It has recently been suggested that the existence of bare strange stars is incompatible with low scale gravity scenarios.  It has been claimed that in such models, high energy neutrinos incident on the surface of a bare strange star would lead to catastrophic black hole growth.  We point out that for the flat large extra dimensional case, the parts of parameter space which give rise to such growth are ruled out by other methods.  We then go on to show in detail how black holes evolve in the the Randall-Sundrum two brane scenario where the extra dimensions are curved.  We find that catastrophic black hole growth does not occur in this situation either.  We also present some general expressions for the growth of five dimensional black holes in dense media.
\end{abstract}
\section{Introduction}
The idea that the geometry of extra dimensions might be responsible for the hierarchy between the scale of electroweak physics and the Planck scale is extremely interesting.  In these models, the mass scale associated with gravity is around a TeV but appears to be much higher due to the small overlap of the extra dimensional graviton wave function with our standard model brane \cite{add,randall}.  
In a gravity theory with $4+d$ space-time dimensions and a fundamental scale $M_F$, one expects black hole production at energy densities higher than $M_F^{4+d}$, so there has been a great deal of interest in the possibility of black hole production at the next generation of super-TeV scale colliders \cite{dimopoulos,cheung,Giddings:2002bu}.
The idea that colliders might produce small black holes is at first alarming, but these black holes are so small that they are expected to evaporate via Hawking evaporation before they are able to interact with their surroundings and grow.

A different situation would arise if the black hole were produced in an extremely dense medium like the interior of a neutron star, as in that case the black hole might interact with another particle before it decays, so that the Hawking evaporation would be balanced by the accretion of matter and the black hole might start to grow.

Production of the initial black hole requires that a nucleon belonging to the star be hit by an incident highly energetic particle such as a cosmic ray or a cosmic neutrino, with an energy of at least a few PeV to reach the threshold of black hole production, $\sqrt{2\ m_N \ E_i} \sim M_{BH} \sim$ few TeV.  According to the hoop conjecture, the cross section for black hole production can be taken to be $\sigma_{BH} = \pi {r_s}^2$ where $r_s$ is the Schwarzschild radius of the centre of mass energy of the incident particle and the target.  Cosmic neutrinos could be a candidate for black hole production since $\sigma_{BH}$ dominates over all the Standard Model neutrino-nucleon interactions for neutrino energies above $\sim100$ PeV \cite{Feng:2001ib}. Ultra High Energy neutrinos are expected to exist (as well as the already observed UHE cosmic rays \cite{AGASA,FLYEYE}), although the current sensitivity of neutrino telescopes does not enable us to detect them \cite{nutelescopes}. The most straightforward mechanisms of production would be via the interaction of UHE cosmic rays with the cosmological microwave background (GZK mechanism \cite{GZK}) and via collisions of accelerated hadrons and photons inside astrophysical objects such as Active Galactic Nuclei. Other, more exotic production processes involving ``hidden sources'' or decay of ultra-heavy relic particles have also been proposed, possibly giving rise to many neutrinos with energies as high as $10^{22-23}$ eV \cite{fluxes}. We prefer however to retain a more conservative estimate of the high energy neutrino flux, essentially based on the assumption that neutrinos are produced by the same cosmologically distributed extra-galactic sources that would be responsible for the observed high energy cosmic rays: the Waxman-Bahcall bound \cite{WB}.  Using this bound, one can deduce the number of neutrinos of energy $E_{min}<E<E_{max}$ falling on a star of radius $R$ per unit time
\begin{equation}
\dot{N} = 800 \ \pi^2 \left ( \frac{R}{1 \ km}\right )^2 \ \left (\frac{1\ {\rm GeV}}{E_{min}} - \frac{1\  {\rm GeV}}{E_{max}}\right )\ \ \   s^{-1}.
\label{waxman}
\end{equation}
This rate would become comparable to the corresponding expression for cosmic rays as the energy increases.  For the surface of a star with a radius of 10 km this rate is of about 40 neutrinos per year with an energy between $10^{20}$ eV and $2\times 10^{20}$ eV, while the current measurements made on Earth, although still quite imprecise, would imply approximately 5 to 20 cosmic rays per year around $10^{20}$ eV.    

A recent paper \cite{learned} has pointed out that such a black hole formed by high energy neutrinos on the outside of a neutron star will not in fact grow since the region in which the black hole first forms is not dense enough for the black holes to interact with more nucleons before it decays.  The same paper also shows that the situation is fundamentally different in the case of strange stars.

It is postulated that the energy per quark in normal (up and down) quark matter may be higher than that in strange quark matter \cite{witten} so it has been hypothesised that one possible end point of stellar evolution is a star entirely composed of up, down and strange quarks \cite{olinto}.  Since it is thought that some strange stars may be `bare' in as much as their density rises from zero to more than nuclear densities in a length scale of order $\sim$ fm \cite{olinto}, strange stars could provide a medium in which TeV scale black holes could be created and then subsequently grow.  One would not expect a strange star to possess spectral lines and also for it to have a different mass-radius relation and cooling rate to a neutron star.   There has recently been a lot of interest in a possible strange star candidate \cite{drake} although it is not clear yet if the identification of this source as a strange star is correct \cite{prasanna}.

The authors of \cite{learned} go on to say that the existence of a bare strange star would place constraints on the number and size of extra dimensions.  This is because for large enough extra dimensions and conservative estimates of high energy neutrino fluxes, the growth of a neutrino-nucleon interaction induced black hole will continue until it consumes the whole of the star.  The constraints obtained in that paper are mainly for the case of $d\le 2$ flat extra dimensions.  

There are many other stronger constraints on the case of 2 extra flat dimensions from astrophysics and cosmology \cite{2con}.  For the case of a single extra dimension which solves the hierarchy problem, it is not possible for the extra dimension to be flat, since this would require physics to be effectively five dimensional at lengths up to solar system scales.  One therefore requires a warped extra dimension as in the model of Randall and Sundrum.  Phenomenologically such theories are difficult to constrain since the graviton Kaluza-Klein mode masses are of order of the fundamental scale $\sim$ TeV \cite{randall}.  This is fundamentally different from the flat extra dimension scenarios where the Kaluza-Klein masses are far below a TeV and can therefore be excited at astrophysical energies.  Given the recent possible detection of strange star candidates, it is interesting to find out if the existence of strange stars would place any constraints upon such scenarios.

In this work we briefly review the 5D Randall-Sundrum model and the black holes that can be formed in this theory.  We then write down some general equations describing the evolution of 5D black holes in dense media.  We show why TeV scale black holes created at the surface of neutron stars do not continue to grow and then describe the growth of black holes in strange stars assuming the existence of a single warped extra dimension.

\section{TeV black holes in Randall-Sundrum\label{sundrum}}
 
In the model of Randall and Sundrum with a compact extra dimension, a large apparent mass hierarchy between gauge and gravitational mass scales is obtained via a warping of the transverse space \cite{randall} (the evolution of black holes in the non-compact Randall Sundrum scenario is studied in \cite{raf}) .  In this study we will assume that there is only one extra dimension although the analysis could easily be extended to $n$ extra warped directions.   The Schwarzschild radius of a black hole of mass $M_B$ in a $4+d$ dimensional flat space-time with a gravitational scale of $M_F$ is given by \cite{perry}\footnote{This definition of the Schwarzschild radius adopts $M_F^{d+2}=(2\pi)^{d}/4\pi G_{d+4}$.} 
\begin{equation}
r_s=\left[\frac{2^{(d+1)}\pi^{(d-3)/2}\Gamma\left(\frac{d+3}{2})\right)}{d+2}\right]^{\frac{1}{1+d}}\frac{1}{M_F}\left(\frac{M_B}{M_F}\right)^{\frac{1}{1+d}}.
\label{flatradius}
\end{equation}
Since the cross section for accretion onto the black hole is therefore proportional to $M_B^{2/(1+d)}$ we will need a higher density medium in order for the black hole to start to grow if $d>1$.  Hence the situation with one extra warped direction is more likely to promote back hole growth.

Using the conventions of the original paper $\cite{randall}$ we write the five dimensional metric
\begin{equation}
ds^2 = e^{- 2 k r_c |\phi|} \eta_{\mu \nu} dx^{\mu} dx^{\nu} + r_c^2 d \phi^2
\end{equation}
where $\phi$ is the coordinate of the orbifold direction ($0<|\phi|<\pi$) and $r_c$ is the size of the compact space.  In the Randall-Sundrum two brane scenario, more than in the flat large extra dimension compactifications, it is really not so clear which mass scale ($10^{19}$ GeV or $1$ TeV) is fundamental and which is derived from the geometry.  We choose to denote the TeV scale $M_F$ and the apparent four dimensional Planck scale $M_{P}$.  Then
\begin{equation}
M_F=M_{P}e^{-\pi k r_c}
\end{equation}
so that we need $k r_c\sim 10$ to solve the hierarchy problem.  The inverse curvature radius of the slice of $AdS_5$ between the branes as viewed from our brane is given by
\begin{equation}
\mu=k e^{-\pi k r_c}.
\end{equation}
Black holes of size $M_F^{-1}\leq r \leq \mu^{-1}$ see an effectively flat 5D space-time \cite{holog} so they obey equation (\ref{flatradius})
\begin{equation}
r_s=0.65\frac{1}{M_F}\left(\frac{M_{B}}{M_F}\right)^{\frac{1}{2}}.
\label{5dhole}
\end{equation}
Since we require that the mass of our black hole is greater than a few times the fundamental scale in order for semi-classical assumptions about the formation and evaporation process to be valid, we now have an expression telling us for which black hole masses we can use the flat space equations
\begin{equation}
M_F \ll M_{B} \lesssim \frac{M_F^3}{\mu^2}.
\end{equation}
In order to accommodate the hierarchy between the scale of electroweak and gravitational physics using this warped geometry, we simply need to ensure that $kr_c\sim 10$, so it appears we can reduce $\mu$ arbitrarily.  However, the further below $M_F$ we take $\mu$ to be, the less natural is the value of $r_c$ required.  Also there is a lower limit on $\mu$ set by the lack of KK mode production at colliders $\cite{oblique}$.

At mass scales such that the radius of the black hole is much larger than the AdS$_5$ radius, full black hole solutions are still out of reach and the behaviour of black hole growth is less clear (although see \cite{Wiseman:2002nn}).  It seems that there are two possibilities for the subsequent behaviour.

The first is that the scattering cross section, and hence the effective size of the black hole, will increase logarithmically \cite{holog,froissart}.  The horizon radius will therefore grow as  
\begin{equation}
r_s\sim\frac{1}{\mu}\ln\left(\frac{\mu^2M_B}{M_F^3}\right)
\label{warped}
\end{equation}
until it fills the space between the branes after which it will continue to grow as a 4D black hole.

Another possibility is that the black hole growth will be suppressed in the radial direction, but will continue along the brane according to the normal equation for a 5D black hole (\ref{5dhole}).  If this occurs then once a black hole has reached the $AdS_5$ radius it will rapidly become entropically favourable for it to split up into an ensemble of many smaller black holes each of which obey the normal flat space relation \cite{suranyi}.

\section{Evolution of 5D black holes in dense media}
If the black hole comes within a distance equal to the Schwarzschild radius of the centre of mass energy of a particle and the black hole itself, the black hole will accrete the particle and continue with a correspondingly larger Schwarzschild radius.

Although classically all of the matter approaching closer than 1.5 times the Schwarzschild radius will be absorbed by the black hole, in \cite{Giddings:2002bu} it was pointed out that much of the energy of a black hole formed with this enhanced cross section will be 'hair', and would be radiated away in a time-scale (for 5 dimensional black holes)
\begin{equation}
\tau_{hair}\sim \left(\frac{M_F}{M_B}\right)^{\frac{3}{2}}\tau_{evaporation}
\end{equation}
Since we only consider situations where the mass of the black hole is at least several times larger than the fundamental scale, we will assume this extra mass is lost on a time-scale much shorter than the black hole lifetime, and we will use the naive geometric cross section $\sigma=\pi r_s^2$.\footnote{Here we neglect the interesting suppression mechanism of Voloshin \cite{voloshin}.  If we were to include this effect, the creation and growth of black holes would be further suppressed}

\subsection{Dense matter with T $\ll$ m}
Consider a black hole moving through a homogeneous medium of particles of mass $m$ and number density $n$ at zero temperature.  
The mean free path of the black hole $\lambda$ before it accretes another particle is given by 
\begin{equation}
\lambda=\frac{1}{n\sigma}=\frac{1}{n \pi r_s^2}
\label{mfp}
\end{equation}
and the rate of increase of mass of the black hole is set by the inverse of the time taken for the BH to cross one mean free path ($\beta=p/E$, $\gamma=E/M_B,E^2=p^2+M_B^2$)
\begin{equation}
\left.\frac{dM_B}{dt}\right|_{acc}=\frac{\beta m}{\lambda}=\beta n \sigma m.
\end{equation} 
Hawking radiation means that the black hole radiates with a temperature given by \cite{hawking}\cite{Giddings:2002bu}
\begin{equation}
T_H=\frac{1+d}{4\pi r_s}
\label{temp}
\end{equation}
and armed with this expression we can use Wien's law to obtain the mass loss of the black hole \cite{dimopoulos}
\begin{equation}
\left.\frac{dM_B}{d\tau}\right|_{evap}=-\frac{\pi^2g_{eff}}{60}A T_H^{4+d}=-\frac{g_{eff}}{960 \pi r_s^2}
\label{evaprate}
\end{equation}
where we use $\tau$ to denote the time coordinate in the rest frame of the BH and $A$ for the horizon area which is given in the appendix.  The parameter $g_{eff}$ is the number of effective relativistic degrees of freedom in the plasma which, since the Hawking temperature is typically below the mass of most of the Kaluza-Klein modes, refers to the standard model degrees of freedom on the brane.  Combining the mass accretion and Hawking evaporation in the lab frame we end up with
\begin{eqnarray}
\frac{dM_B}{dt}&=&\left.\frac{dM_B}{dt}\right|_{acc}+\frac{1}{\gamma}\left.\frac{dM_B}{d\tau}\right|_{evap}\nonumber\\
&=&\pi r_s^2\beta m\left\{n-\frac{g_{eff}}{960\pi^2 m}\frac{M_B}{pr_s^4}\right\}\label{ncrit}
\end{eqnarray}
This equation shows us straight away that for each black hole mass $M_B$ and momentum $p$ there is a critical number density of particles for which the rate of mass gained through accretion will be greater than the rate of mass lost through evaporation.

One might expect that solution of this differential equation would lead to a full description of the behaviour of the black hole.  If this were so a rapidly moving black hole would accrete matter until it became stationary.  This would occur when its gamma factor goes down to close to 1 so the mass of the final black hole at rest would be of the same order of its initial momentum.  However, this is not the case since the momentum $p$ does not remain constant.  

As the black hole accretes matter $p$ is indeed conserved but each time the black hole evaporates a particle of mass $m_{out}$ the black hole will lose a fraction of its momentum $p$ such that $\Delta p/p=-m_{out}/M_B$.  This is associated with the fact that the wavelengths of the quanta emitted via Hawking radiation are greater than the radius of the black hole so that the process is effectively s-wave emission in the black hole rest-frame.  We therefore have an additional equation which governs the evolution of $p$
\begin{equation}
\frac{dp}{dt}=\frac{p}{M_B}\frac{1}{\gamma}\left.\frac{dM_B}{d\tau}\right|_{evap}
\label{mom}
\end{equation}
which we have to solve simultaneously with equation (\ref{ncrit}) in order to obtain the evolution of the black hole.

\subsection{Application to Outer Neutron Star Crust}
The outside $\sim 300$ metres of a neutron star consists of a degenerate electron gas with the nucleons becoming increasingly neutron rich as one moves inwards to higher densities \cite{pethick}.  At a nucleon number density of about $2\times 10^{-4}{\rm fm}^{-3}$, it is energetically favourable for the neutrons to occupy continuum states and they begin to drip out of the nuclei.

In the outer parts of the neutron star crust, the pressure energy density is much less than the mass energy density, and the total mass of the crust is only $1\%$ of the neutron star mass.  The Tolman-Oppenheimer-Volkoff equation \cite{tov} is the general relativistic equation for hydrostatic equilibrium
\begin{equation}
\frac{dP}{dr}=-\frac{\left(\rho+P\right)G\left(m(r)+4\pi r^3P\right)\Lambda(r)}{r^2}
\end{equation}
where $\Lambda(r)=(1-2Gm(r)/r)^{-1}$.  In the crust of the neutron star $\rho(r_d)\sim 2\times 10^{30} {\rm eV}^4$ and $P(r_d)\sim 8\times 10^{27} {\rm eV}^4$ so $P/\rho<1\%$ \cite{pethick}.  Also, in the crust $m(r)$ varies from the total mass of the neutron star $M_{*}$ by at most a few percent, so we can write
\begin{equation}
\frac{dP}{dr}\simeq-\frac{GM_{*}\Lambda(R_{*})}{R_{*}^2}\rho
\end{equation}
Here $R_{*}$ is the radius of the neutron star which we take to be 10 km and we set $M_{*}=1.4M_\odot$.  We denote the radius at which neutron drip occurs as $r_d$ and assume that $\rho=m_n n$ where n is the number density of nucleons.  The electron density $n_e$ is relativistic so that\footnote{This is in reasonable agreement with the values obtained in the more detailed analysis of \cite{pethick}.} $P=(\pi/2)(3/8\pi)^{1/3}n_e^{4/3}$ and if we make the simplifying assumption that $n_e=n_p=n$ we can write
\begin{equation}
n(r_d)^{\frac{1}{3}}-n(r)^{\frac{1}{3}}=\frac{GM_{*}\Lambda(R_{*})}{4R_{*}^2}m_n(r-r_d)
\end{equation}
with $n(r_d)\sim10^{22} {\rm eV}^3$.  An incident neutrino with momentum $10^{19} {\rm eV}$ will collide with a neutron to create a black hole of mass $M_{B}\sim 10^{14} {\rm eV}$.  Using the expression for the mean free path (\ref{mfp}) one finds that such a neutrino moving through this outer crustal region will have a mean free path of $\lambda\sim ({\rm eV}^{-1})$ so will not penetrate any deeper into the star before becoming a black hole.  The critical density for the growth of a black hole with such a mass and momentum can be calculated from equation ($\ref{ncrit}$) and is $n\sim 10^{29}{\rm eV}^{3}$ so the black hole will not grow in this outer region of the star where it is created.

The medium will therefore become optically thick due to the neutrino-nucleon black hole production cross section, however, this will happen at a depth much smaller than that where the surrounding density of matter is high enough for the resulting black hole to accrete more matter than it will evaporate.  The black hole will therefore decay thermally into standard model particles which will join the surrounding star.  For very large incident neutrino energies, some of the decay products may be able to produce secondary black holes, but these black holes will also decay in a time scale much smaller than the time required for them to travel into the depths of the neutron star where they may be able to grow.  Our conclusion for neutron stars is therefore the same as the authors of $\cite{learned}$.

\subsection{Radiation with T $\gg$ m}

\begin{figure}
\begin{center}
\includegraphics[height=8cm,width=13cm]{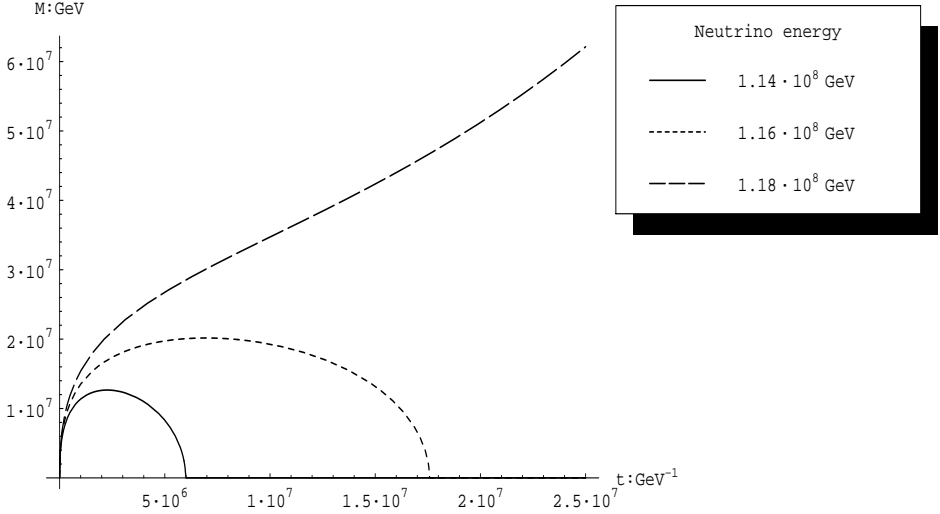}
\end{center} 
\caption{\label{flatm}Evolution of black hole mass vs. time when the black hole is much smaller than $\mu^{-1}$ for different neutrino energies for $T=1$ GeV, $M_F=1$ TeV and $M_{Binitial}=5$ TeV.  This growth threshold is not valid for black holes of size $r_s>\mu^{-1}$.}
\end{figure}

\begin{figure}
\begin{center}
\includegraphics[height=8cm,width=13cm]{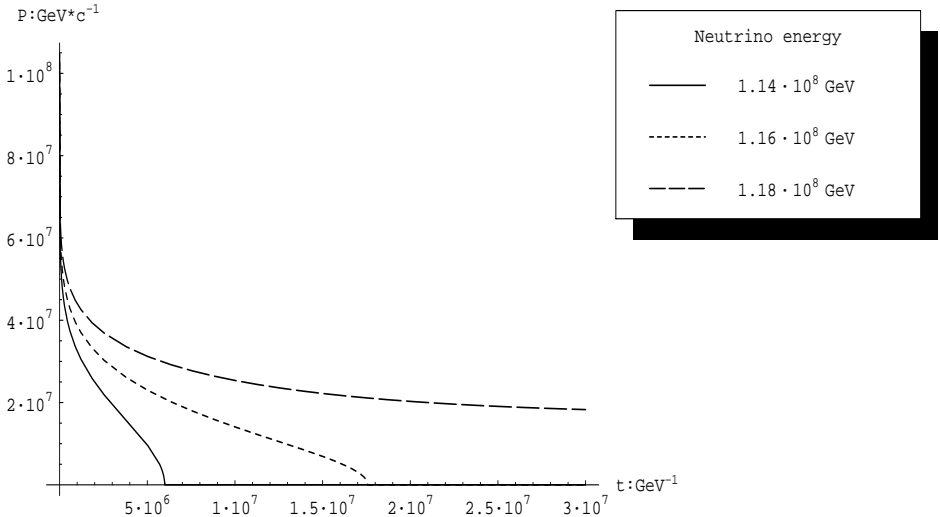}
\end{center} 
\caption{\label{flatp}Evolution of black hole momentum vs. time for $r_s<\mu^{-1}$ for different neutrino energies, $T=1$ GeV, $M_F=1$ TeV and $M_{Binitial}=5$ TeV.}
\end{figure}

Now if we consider the situation where the black hole is moving through a medium of relativistic particles at temperature $T_{bath}$, it becomes more convenient to work in the rest frame of the black hole.  Following \cite{learned}, the effective temperature of the medium in the black hole frame becomes

\begin{equation}
T_{eff}=T_{bath}\sqrt{\gamma}\left(1+\frac{\beta^2}{3}\right)^{\frac{1}{4}}=T_{bath}\left(1+\frac{4}{3}\frac{p^2}{M_B^2}\right)^{\frac{1}{4}}.
\end{equation}
If one assumes that all the effectively light degrees of freedom are constrained to lie on the brane, the mass of a 5D black hole will evolve in its own rest frame as
\begin{equation}
\frac{dM_{B}}{d\tau}=\frac{\pi^3 g_{eff}}{15}r_s^2\left(T_{eff}^4-\frac{1}{4(2\pi)^4 r_s^4}\right)
\label{tgrow}
\end{equation}
and the evolution of momentum will still be given by equation (\ref{mom}).  In figures \ref{flatm} and \ref{flatp} we show the evolution of the mass and momentum of black holes formed by neutrinos with various momenta propogating in a medium of temperature $T=1$ GeV.  These figures show the evolution in the region where $r_s<\mu^{-1}$ so that equation (\ref{flatradius}) gives the cross section.  As we will see in the next section, when $r_s>\mu^{-1}$ it becomes much more difficult to obtain growth.  Put another way, black holes that are able to grow during the flat regime often cannot continue to grow once they become larger than the curvature of the compact space.  The examples in the figures correspond to values of $\mu$ that have been ruled out at accelerators.

\subsection{Application to Strange Star Interiors}
The surface of a strange star is expected to have a thin shell of electrons with thickness of the order of a few hundred fm \cite{olinto}.  The Coulombic repulsion of this shell will not be able to stop matter falling freely onto the star from infinity, however if the accretion onto the star is in the form of a fluid the incoming matter may lose energy via the normal accretion processes.  A crust would then build up on the exterior of the strange star which might create the same barrier to black hole growth as the outer crust of a neutron star.  However, if strange stars exist, one would expect at least some of them to exist outside binary systems and therefore possess surface density profiles very close to step functions.  It therefore seems possible that such stars would indeed provide a suitable medium for the growth of TeV scale black holes.  The interior of a strange star consists of quarks with a temperature of about $T_{bath}\sim 1$ GeV \cite{olinto}.  Since we have shown that neutrinos of sufficiently high energy will create black holes that grow in a medium of $T=1$ GeV, we need to consider what will happen once the size of the black hole reaches $\mu^{-1}$.
\subsubsection{Black Holes with $r>\mu^{-1}$}
As discussed earlier, once such a black hole becomes as large as the $AdS_5$ curvature radius $\mu^{-1}$, there are two possibilities as to its future evolution.
\begin{itemize}
\item The first is that presented in \cite{suranyi} where the black hole splits up into many smaller black holes once it grows out to the $AdS_5$ radius.  In this scenario, there is a minimum evaporation temperature for 5D black holes, since if they grow larger than $\mu^{-1}$ they will decay into smaller holes with higher temperatures.  As the holes grow and split up, they will also gain mass whilst losing momentum via radiation and ultimately the following criterion applies to such a system of black holes: If they are to have any chance of growth, the rest frame temperature of the medium must be higher than the minimum black hole temperature ($\sim \mu$).
Using equation (\ref{tgrow}) to obtain a more precise estimate we find we require a value of $\mu < 20$ GeV in order for a 5D Randall-Sundrum black hole to have any chance of growing in this medium.  This region of parameter space has already been ruled out by the fact that Kaluza-Klein modes have not lead to loop corrections of the oblique parameters in electroweak interactions \cite{oblique}.

\item The second scenario for growth of black holes is our extrapolation of the scenario described in \cite{froissart} where the size of the black hole will only increase logarithmically until enough mass has been added so that the black hole fills the bulk.  Once the black hole fills the bulk, its subsequent evolution will obey the normal 4D mass radius relation $M_B\sim M_P^2 r_s$.  This will typically occur at a mass more than $30$ orders of magnitude higher than the mass at which the radius reaches $\mu^{-1}$. 
The fact that the radius of the hole now only increases slowly with mass means that the temperature remains high as the mass increases.  This would suggest that once black holes are big enough to feel to the curvature of the compact space, their growth will be suppressed since the radius fails to increase as rapidly with mass after that point.  This is exactly what we find in our numerical analysis.
\end{itemize}
\subsubsection{Distance Travelled by Black Hole}
We also need to check whether or not the black hole actually stays inside the star, as black holes that have sufficient momentum to grow may quickly traverse the star and exit.  The distance travelled by the black holes is easily calculated via
\begin{equation}
{\rm distance}=\int \beta dt = \int \frac{p(t)}{E(t)}dt=\int \frac{p(\tau)}{E(\tau)}\gamma(\tau)d\tau=\int \frac{p(\tau)}{M_B(\tau)}d\tau.
\end{equation}
We have assumed a black hole travelling more than 10 km will exit the star, but we will see that the exact figure is not critical.

If a black hole should exit the star, it is then  necessary to calculate the velocity of the black hole at that time and compare it with the escape velocity of the star given by
\begin{equation}
\beta_{escape}\sim\sqrt{\frac{2GM_*}{R_{*}}}=0.94
\end{equation}
where we have assumed the star has a mass $M_*$ equal to the sun $\sim 10^{57}$ GeV and a radius $R_{*}$ of 10 km.  Again, we will see that the exact values are not critical.  Black holes which exit the star with velocity $\beta_{exit}>\beta_{escape}$ will simply escape to infinity.  For the case of black holes which exit the star with $\beta_{exit}<\beta_{escape}$, we need to check if they will evaporate away completely before they fall back onto the surface of the strange star and start to accrete again.  Therefore it is necessary to compute the decay time 
\begin{eqnarray}
t_{decay}&=&\frac{960\pi}{g_{eff}}\frac{1}{\mu}\int_{0}^{M_B}\ln^2\left(\frac{\mu^2 M}{M_F^3}\right) dM\nonumber\\
&=&\frac{960\pi}{g_{eff}}\frac{M_B}{\mu^2}\left[\ln^2\left(\frac{\mu^2 M_B}{M_F^3}\right)-2\ln\left(\frac{\mu^2 M_B}{M_F^3}\right)+2\right]
\end{eqnarray}
which uses the expression for evaporation (\ref{evaprate}) and the equation fdor the radius (\ref{warped}) since the black holes in our simulations have $r\gg\mu^{-1}$.  We must then compare this with the time taken for the black hole to fall back onto the surface of the star 
\begin{eqnarray}
t_{return}=2\int_{R_{*}}^{r_{turn}}{\frac{1}{\beta(r)}}dr&=&2\int_{R_{*}}^{r_{turn}}\left[\beta_{exit}^2+2GM_*\left(\frac{1}{r}-\frac{1}{R_{*}}\right)\right]^{-\frac{1}{2}} dr \nonumber\\
&=&2\frac{\beta_{exit}R_{*}^2}{GM_*}
\end{eqnarray}
where the maximum distance of the black hole from the star is given by
\begin{equation}
r_{turn}=\left(\frac{\beta_{exit}^2}{2GM_*}+\frac{1}{R_{*}}\right)^{-1}.
\end{equation}
\subsubsection{Numerical Analysis}
The longest known strange star candidate was observed only about 10 years ago \cite{walter} and using the Waxman-Bahcall bound (\ref{waxman}) we can estimate the maximum energy neutrino that may have been incident on that object since its discovery.  The answer is $2\times 10^{23}$ eV which is very high compared to observed cosmic ray showers.  However, since the growth of a black hole is more likely for higher energy incoming particles, adoption of this energy makes our constraint stronger.

\begin{figure}
\begin{center}
\includegraphics[height=8cm,width=13cm]{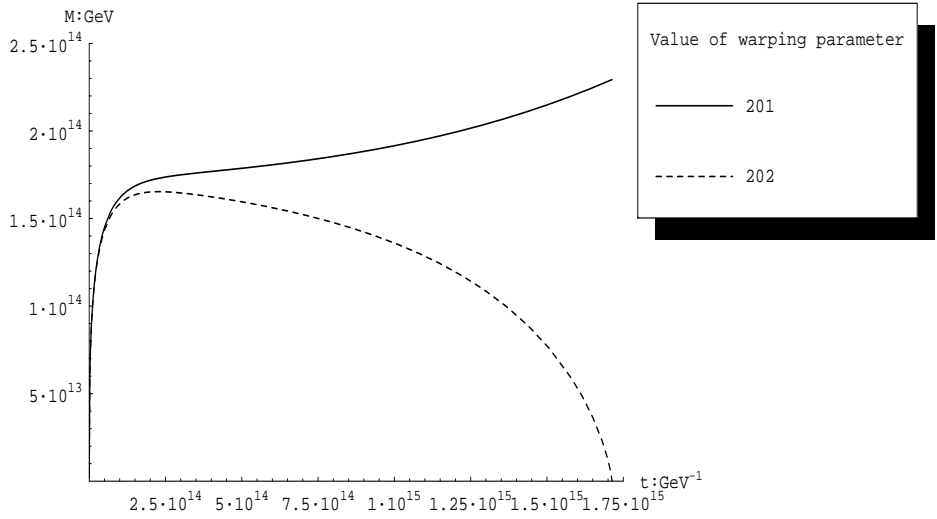}
\end{center} 
\caption{\label{202}Evolution of black hole for $\mu=$201 and 202 GeV showing that for $M_{F}=$1 TeV, $M_{Binitial}=5$ TeV and $E_{\nu}=2\times 10^{23}$ eV this value of $\mu$ is the critical one for growth.  Note that such a black hole will leave the star after a time $t=8\times 10^9 {\rm GeV}^{-1}$ with a mass much smaller than the total mass of the star ($\simeq 10^{57}$ GeV).}
\end{figure}

The results of our investigations are summarised in table \ref{restable}.  We find that {\it no} black holes formed at these energies grow quickly enough to remain inside the star.  For very high values of $\mu$ ($\mu\ge$ 202 GeV,see figure \ref{202}) black holes are not able to continue their initial growth and start to lose mass faster than they accrete.  These black holes exit the star and escape to infinity (although they will of course decay rather quickly once they are outside the star).  For lower values of $\mu$, the black holes are able to grow but exit the star long before they engulf it.  These black holes also escape with a velocity higher than the escape velocity.  For the lowest values of $\mu$ the black holes grow more quickly due to the fact that they spend a longer period of time in the flat regime where their cross sections grow rapidly.  They therefore exit the star with lower velocities, but still decay before they are able to fall back onto the star.

\begin{table*}
\begin{center}
\caption[]{\label{restable}Summary of evolution of black holes created by a neutrino of energy $2\times 10^{23}$ eV for different values of the curvature parameter $\mu$.  No such black holes remain in the star.  We assume a strange star of radius 10km and temperature 1 GeV, and an initial mass for the black hole of 5 TeV.}
\begin{tabular}{|c|c|}\hline
202 GeV$\leq\mu$&black hole does not grow\\
&$\beta_{exit}>\beta_{escape}$\\ \hline
50 GeV$\leq\mu\leq$201 GeV&black hole grows\\
&$\beta_{exit}>\beta_{escape}$\\ \hline
$\mu \leq$ 49 GeV&black hole grows\\
&$\beta_{exit}<\beta_{escape}$\\
  &${\rm t}_{return}\gg {\rm t}_{decay}$\\ \hline	
\end{tabular}
\end{center}
\end{table*}

To see this, we remember that the lowest value of $\mu$ that is not ruled out at collider experiments is $\mu\sim20$ GeV.  For this value of $\mu$, a black hole created by a neutrino with the initial momentum described earlier leaves the star with a velocity of $\beta_{exit}= 0.6$ which corresponds to a return time of $t_{return}= 6\times 10^{19}{\rm GeV}^{-1}$.  The decay time for this black hole is $t_{decay}= 1\times 10^{16}{\rm GeV}^{-1}$.  In order to find a black hole that could fall back onto the star before it decays we would therefore have to consider values of $\mu$ less than the permitted experimental lower limit.  Thus black holes formed from high energy neutrinos will not grow to engulf strange stars.

\section{Conclusion}

In this work we investigated the possibility that the Randall-Sundrum 2-brane model of TeV scale gravity is incompatible with the existence of bare strange stars due to the growth of black holes seeded by high energy neutrinos.  We pointed out that the growth of such black holes is suppressed in these models when they reach the radius corresponding to the curvature radius of the $AdS_5$ in between the branes.  We performed detailed simulations to see if black holes would grow and engulf the star.  We saturated the Waxman-Bahcall bound to find the highest energy neutrino that one could expect to have hit the oldest known strange star candidate \cite{walter} in the time since it has been discovered.  In doing so, we found that the regions of parameter space where growth of such holes may be permitted has already been ruled out by accelerator experiments.

In the paper of \cite{learned} it was shown that catastrophic black hole growth in strange stars could only work if there were 1 or 2 large extra dimensions.  The case of 2 large flat extra dimensions has already been tightly constrained by astrophysical and cosmological constraints \cite{2con}.  In this paper we have ruled out the possibility of black hole growth in the Randall-Sundrum model.  We therefore do not expect strange stars to collapse due to TeV gravity black hole seeding.

We have also presented formalisms for the growth of 5D black holes in dense media which might be of interest in other areas such as cosmology and hadronic collider physics.

\section*{Acknowledgements}
We would like to thank Nicolas Borghini, Robert Brout, Dominic Clancy, Jean-Marie Frere, Steve Giddings, James Gray, Raf Guedens, Scott Thomas and Toby Wiseman for interesting discussions.  We would also like to thank the referee for his/her invaluable comments.  MF is funded by an IISN grant and the IUAP program of the Belgian Federal Government.

\section*{Area of 2+d Spheres}
The surface area of sphere in $3+d$ space dimensions is given by the expression
\begin{equation}
A=\frac{(3+d)\pi^{\frac{3+d}{2}}}{[(3+d)/2]!}r_s^{2+d}
\end{equation}
where for half integer factorials we use
\begin{equation}
(1/2+n)!=\sqrt{\pi}\frac{(2n+2)!}{(n+1)! 4^{n+1}}
\end{equation}

\end{document}